%Paper: hep-th/9501091
%From: klimcik@surya11.cern.ch (Ctirad KLIMCIK)
%Date: Fri, 20 Jan 95 11:59:31 +0100

\magnification=1200

\
\rightline{hep-th/9501091}
\vskip3cm

\centerline
{\bf `NULL GAUGED' WZNW THEORIES}
\centerline{\bf AND TODA-LIKE $\sigma$-MODELS}
\vskip2cm

\centerline{C. Klim\v c\'\i k \footnote{$^{*}$}{\baselineskip5pt
e-mail: klimcik@surya20.cern.ch} }
\vskip1cm
\centerline {\it  Theory Division, CERN,   }
\centerline {\it   CH-1211 Geneva 23, Switzerland}
\vskip1cm

\centerline{\bf Abstract}

\vskip1cm
\noindent
A new method of gauging of WZNW models is presented, leading to a class
of exact string solutions with a target space metric of
Minkowskian signature. The corresponding models may be interpreted
as $\sigma$-model analogues of the Toda field theories.

\vskip2cm

Talk given at the $28^{th}$ International Symposium on the Theory of
Elementary Particles, Wendisch-Rietz, 30 August - 3 September 1994
\vfill\eject

%%%%%%%%%%%%%%%
%%%%%%%%%%%%%%%%%%%%%%%%%%%%%%%%%%%%%%%%%%%%%%%%%%%%%%%%%%%%
\input harvmac
%%%%%%%%%%%%%%%%%%%%%%%%%%%%%%%%%%%%%%%%%%%%%%%%%%%%%%%%%

\def  \TF {\tilde F}
\def \tp {{\tilde \p}}

\def\bd {{\bar \del}}\def \ra {\rightarrow}

\def \EE {{\bar {\cal  E}}}
\def \ll {{\bar \l}}
\def \J {{\bar J}}
\def \M {{\cal M}}
\def \ao {{\a_1}}
\def \am  {{-\a_1}}

\def \tu {\tilde u }
\def \tv {\tilde v}
\def \sms {sigma models \ }
%%%%%%%%%%%%%%%%%%%%%
\def \ca
\def \P {\Phi}

 \def \k1 {{1\over
k}} \def \bh { {\bar h} } \def \ov { \over }
  \def \B { { \bar B }}

\def \ra {\rightarrow}

\def \a {\alpha}
\def \b {\beta}

\def \Tr {{\ \rm Tr \ }}

\def \ln {{\rm \ ln \  }}
\def \det {{ \rm det \ }}

\def \l {\lambda}\baselineskip8pt

\vfill\eject

\def \1p {{1\over  \pi }}
\def \2p {{{1\over  2\pi }}}
\def \4p {{ {1\over 4 \pi }}}
\def \8p {{{1\over 8 \pi }}}
\def \P^* { P^{\dag } }
\def \p {\phi}
\def \M {{\cal M}}
\def \m {\mu }
\def  \n {\nu}
\def \ep {\epsilon}
\def\g {\gamma}
\def \r {\rho}
\def \k {\kappa }
\def \d {\delta}

\def \s {\sigma}

\def \eq#1 {\eqno {(#1)}}
%%%%%%%%%%%%%%%%%%%%%%%%%%%%%%%%%%%%%%%%%%%%
\def \sm {sigma model\ }\def \B  {{ \tilde B }}

\def \bd  {{ \bar \del }}

\def \M {{\cal M}}

\def \B  {{ \tilde B }}

\def \bd  { \bar \del }

\def \ov {\over }

\def \A  { {\bar A} }

\def \bu {{\bar w}}

%%%%%%%%%%%%%%%%%%%%%%%%%%%%%%%%%%%

\def \p {\phi}
\def \ep {\epsilon}
\def \s {\sigma}

\def \r {\rho}
\def \d {\delta}
\def \l {\lambda}
\def \m {\mu}
\def \g {\gamma}
\def  \n {\nu}

\def \B {{\bar B}}

\def \rank {{\rm rank\ }}
\def \J {\bar J }
\def \P {\Phi }
%%%%%%%%%%%%%%%%%%%%%%%%%%%%%%%%%%%%%%%%%%%%%%

\noblackbox
\overfullrule=0pt
\baselineskip 20pt plus 2pt minus 2pt

%%%%%%%%%%%%%%%%%%%%%%%%%%%%%%%%%%%%%%%%%%%%%%%%%%%%%%%%%%%%%%%
%%%%%%%%%%%%%%%%%%%%%%%%%%%%%%%%%%%%%%%%%%%%
\def\np {  Nucl. Phys. }
\def \pl { Phys. Lett. }
\def \mpl { Mod. Phys. Lett. }
\def \prl { Phys. Rev. Lett. }
\def \pr  { Phys. Rev. }

\def \cmp { Commun. Math. Phys. }
\def \ijmp { Int. J. Mod. Phys. }

%%%%%%%%%%%%%%%%%%%%%%%%%%%%%%%%%%%%%%%%%
%%%%%%%%%%%%%%%%%%%%%%%%%%%%%%%%
\lref \anton { I. Antoniadis, C. Bachas, J. Ellis and D.V. Nanopoulos,
\pl B211(1988)393. }
\lref \horav{P. Ho\v rava, \pl B278(1992)101.   }
\lref\gersh {  D. Gershon,  \pr D49(1194)999;  hep-th/9210160,
9311122. }

\lref \bcr {K. Bardakci, M. Crescimanno and E. Rabinovici, \np
B344(1990)344. }
\lref \wiit { E. Witten, \pr D44(1991)314.}
\lref \gwz {      K. Bardakci, E. Rabinovici and
B. S\"aring, \np B299(1988)157;
 K. Gawedzki and A. Kupiainen, \pl B215(1988)119;
\np B320(1989)625. }

\lref \karabali { D. Karabali, Q-Han Park, H.J.
Schnitzer and
Z. Yang, \pl B216(1989)307;  D. Karabali and H.J. Schnitzer, \np B329(1990)649.
}
\lref \polwig { A.M. Polyakov and P.B. Wiegman,  \pl
B141(1984)223.  }

 \lref \kumar {A. Kumar,
\pl B293(1992)49; D. Gershon, preprint TAUP-2005-92.}
\lref  \hussen {  S. Hussan and A. Sen,  \np B405(1993)143. }
\lref \kiri {E. Kiritsis, \np B405(1993)109. }
\lref \alv { E. Alvarez, L. Alvarez-Gaum\'e, J. Barb\'on and Y. Lozano,
preprint CERN-TH.6991/93.}

\lref  \rocver { M. Ro\v cek and E. Verlinde, \np B373(1992)630.}

\lref \brink{ H.W. Brinkmann, Math. Ann. 94(1925)119.}
\lref \guv {R. G\"uven, Phys. Lett. B191(1987)275.}

\lref \amkl { D. Amati and C. Klim\v c\'\i k, \pl B219(1989)443.}

\lref \hor { G. Horowitz and A.R. Steif, Phys.Rev.Lett. 64(1990)260 ;
Phys.Rev. D42(1990)1950.}

\lref \horr {G. Horowitz, in: {\it Proceedings
of  Strings '90},
College Station, Texas, March 1990 (World Scientific,1991).}

\lref \rudd { R.E. Rudd, \np B352(1991)489 .}

\lref \tsnul { A.A. Tseytlin, \np B390(1993)153.}
\lref \tsnull { A.A. Tseytlin, \pl B288(1992)279; \pr D47(1993)3421.}

\lref \dunu { G. Horowitz and A.A. Steif, \pl B250(1990)49;
 E. Smith and J. Polchinski, \pl B263(1991)59;}

\lref\horhorst  {J. Horne, G. Horowitz and A. Steif, \prl 68(1992)568;
G. Horowitz, in:  {\it  Proc. of the 1992 Trieste Spring School on String
theory and Quantum Gravity},
preprint UCSBTH-92-32, hep-th/9210119.}
\lref \horwel {
 G. Horowitz and D.L. Welch, \prl 71(1993)328.
}
\lref \klitse{C. Klim\v c\'\i k and A.A. Tseytlin, \np B424(1994)71.}
\lref \horotse{G.T. Horowitz and A.A. Tseytlin, \pr D50(1994)5204.}

\lref \sfetsos { K. Sfetsos, preprint USC-93/HEP-S1, hep-th/9305074.}
 \lref \busc { T.H. Buscher, \pl B194(1987)59 ; \pl B201(1988)466.}
\lref \pan  { J. Panvel, \pl B284(1992)50. }

\lref \mye {  R. Myers, \pl B199(1987)371;
    I. Antoniadis, C. Bachas, J. Ellis, D. Nanopoulos,
\pl B211(1988)393;
 \np B328(1989)115. }
\lref \givpassq { A. Giveon and A. Pasquinucci, \pl B294(1992)162. }
\lref \GK {A. Giveon and E. Kiritsis,  preprint CERN-TH.6816/93,
RIP-149-93.}
\lref \givroc {A. Giveon and M. Ro\v{c}ek, Nucl. Phys. B380(1992)128.}
\lref \giv {  A. Giveon, Mod. Phys. Lett. A6(1991)2843;  R. Dijkgraaf, H.
Verlinde and
E. Verlinde, \np B371(1992)269; I. Bars, preprint USC-91-HEP-B3;
 E. Kiritsis, \mpl A6(1991)2871. }
\lref \dvv {  R. Dijkgraaf, H.
Verlinde and
E. Verlinde, \np B371(1992)269. }

\lref \GRV {A. Giveon, E. Rabinovici and G. Veneziano, Nucl. Phys.
B322(1989)167;
A. Shapere and F. Wilczek, \np B320(1989)669.}
\lref \GMR {A. Giveon, N. Malkin and E. Rabinovici, Phys. Lett. B238(1990)57.}
\lref \Ve    { K.M. Meissner and G. Veneziano, \pl B267(1991)33;
M. Gasperini, J. Maharana and G. Veneziano, \pl B272(1991)277; \pl
B296(1992)51.}
\lref \sen {A. Sen,  \pl  B271(1991)295; \pl B274(1991)34.}

\lref \VV    {  G. Veneziano, \pl B265(1991)287.}

\lref  \horne { J.H. Horne and G.T. Horowitz, \np B368(1992)444.}

\lref \koki { K. Kounnas and  E. Kiritsis, preprint CERN-TH.7059/93;
hep-th/9310202. }
\lref \napwit { C. Nappi and E. Witten, \prl 71(1993)3751.}

\lref \napwi { C. Nappi and E. Witten, \pl B293(1992)309.}

\lref \tsmpl {A.A. Tseytlin, \mpl A6(1991)1721.}

\lref \tsbh {A.A. Tseytlin,  preprint CERN-TH.6970/93; hep-th/9308042.}

\lref \nsw {  K.S. Narain, M.H. Sarmadi and E. Witten, \np B279(1987)369. }
\lref \call {C.G. Callan, D. Friedan, E. Martinec and M.J. Perry,
Nucl. Phys. B262 (1985)593 ;
E. S. Fradkin and A. A. Tseytlin, \np B261(1985)1;
  A.A. Tseytlin, \pl B178(1986)349.}

\lref \STT {K. Sfetsos and A.A. Tseytlin, preprint CERN-TH.6969/93,
hep-th/9310159.}
\lref \givkir {A. Giveon and E. Kiritsis,  preprint CERN-TH.6816/93;
hep-th/9303016. }

\lref \cec { S. Cecotti, S. Ferrara and L. Girardello, \np B308(1988)436. }

\lref \tsdu { A.A. Tseytlin, \np B350(1991)395.  }

\lref \wit {E. Witten, Commun. Math. Phys. 92(1984)455 ;
E. Braaten, T.L. Curtright and C.K. Zachos, \np B260(1985)630.}

\lref \sftss{ K. Sfetsos and A.A. Tseytlin, unpublished (1994). }
\lref \witte { E. Witten, \cmp 144(1992)189.}
\lref \koulust { E. Kiritsis, C. Kounnas and D. L\"ust, preprint
CERN-TH.6975/93,
hep-th/9308124.}
\lref \kou { C. Kounnas, preprint CERN-TH.6799/93,
hep-th/9304102;
I. Antoniadis, S. Ferrara and C. Kounnas, preprint CERN-TH.7148/94,
hep-th/9402073.}

\lref \oli { D. I. Olive, E. Rabinovici and A. Schwimmer, \pl B321(1994)361.}
\lref \sfet {K. Sfetsos, \pl B324(1994)335; preprints  THU- 93/31, 94/01;
hep-th/9311093,
 9402031.}
\lref \moham { N. Mohammedi, \pl B325(1994)371.}

\lref\ginspqu   { P. Ginsparg and F. Quevedo, \np B385(1992)527.}
\lref\kounn { C. Kounnas and D. L\" ust, \pl B289(1992)56.}
\lref \gidd { S.B. Giddings, J. Polchinski
and A. Strominger, \pr D48(1993)5784.}

\lref  \guv {
R. G\"uven, Phys. Lett. B191(1987)275.}

\lref \amkl { D. Amati and C. Klim\v c\'\i k, \pl B219(1989)443.}
\lref  \hst { G. Horowitz and A.R. Steif, Phys.Rev.Lett. 64(1990)260;
Phys.Rev. D42(1990)1950.}
\lref \petrov { A.Z. Petrov, Einstein Spaces (Pergamon, N.Y. , 1969).}
\lref \figue {J.M. Figueroa-O'Farrill and S. Stanciu, preprint QMW-PH-94-2,
 hep-th/9402035.}
\lref \ira { M. Alimohammadi, F. Ardalan and H. Arfaei, preprint BONN-HE-93-12,
 hep-th/9304024.}
\lref \ind { A. Kumar and S. Mahapatra, preprint IP/BBSR/94-02,
hep-th/9401098.}
\lref \oraif {P. Forg\'acs, A. Wipf, J. Balog, L. Feh\'er and L.
O'Raifeartaigh,
\pl B227(1989)214;  J. Balog, L. Feh\'er,  L. O'Raifeartaigh,
 P. Forg\'acs
and A. Wipf,  Ann. Phys. 203(1990)76;
L. O'Raifeartaigh,
 P. Ruelle and I. Tsutsui, \pl B258(1991)359.}
\lref \sen {A. Sen, \pl B274(1992)34.}

\lref \witt { E. Witten, \cmp 92(1984)455.}
\lref \gwzw  { P. Di Vecchia and P. Rossi, \pl  B140(1984)344;
 P. Di Vecchia, B. Durhuus  and J. Petersen, \pl  B144(1984)245.}

\lref\barsf { I. Bars and K. Sfetsos, \pl B277(1992)269;
 \pr D46(1992)4495, 4510;
I. Bars, preprint USC-93/HEP-B3,  hep-th/9309042. }
\lref\bs { I. Bars and K. Sfetsos,  \pr D48(1993)844. }
\lref \ts {A.A.Tseytlin, \np B399(1993)601; \np B411(1994)509.}
\lref \arfa {H. Arfaei and N. Mohammedi, preprint BONN-HE-93-42,
hep-th/9310169.}

\lref \jack { I. Jack and J. Panvel, preprint LTH-304,  hep-th/9302077.}
\lref \sfts {K. Sfetsos and A.A. Tseytlin, \np B415(1994)116,
hep-th/9308018.}
\lref \jackk { I. Jack, D.R.T. Jones and
 J. Panvel, preprint  LTH-315, hep-th/9308080.}
\lref \tye { S.-W. Chung and S.-H. Tye, \pr D47(1993)4546.}
\lref \giveon { M. Ro\v cek and E. Verlinde, \np B373(1992)630;
 A. Giveon, M. Porrati and E. Rabinovici, preprint RI-1-94,  hep-th/9401139. }

\lref\bakas  {I. Bakas, preprint CERN-TH.7144/94,   hep-th/9402016. }

\lref \hum {J.E. Humphreys, Introduction to Lie Algebras and Representation
Theory
 (Springer, New York 1972) }

\lref \baru { A. Barut and R. Raczka, ``Theory of Group Representations and
Applications" (PWN, Warszawa 1980). }
\lref \zhel {D. Zhelobenko and A. Shtern,  Representations of Lie Groups
(Nauka, Moscow 1983). }

\lref \bal { J. Balog, L. O'Raifeartaigh,  P. For\' gacs
and A. Wipf,  \pl B325(1989)225.
 }
\lref  \barn { I. Bars and D. Nemeschansky, \np B348(1991)89.}
\lref \olive{D.I. Olive, ``Lectures on gauge theories and Lie algebras",
Imperial College preprint (1982).}
\lref \schts{A.S. Schwarz and A.A. Tseytlin, \np B399(1993)691.    }
\lref  \mans { P. Mansfield, \np B222(1983)419.}
\lref \tsey { A.A. Tseytlin, \pl B241(1990)233.}
\lref \ger {A. Gerasimov, A. Morozov, M. Olshanetsky, A. Marshakov and S.
Shatashvili, \ijmp
A5(1990)2495. }
\lref \gervais {J.-L. Gervais and M.V. Saveliev, \pl B286(1992)271.   }
\lref \bil {A. Bilal and J.-L. Gervais, \pl B206(1988)412; \np B318(1989)579. }
\lref \lez {A.N. Leznov and M.V. Saveliev, \cmp 74(1980)111;  89(1983)59.}
\lref \wip {L. O'Raifeartaigh
and A. Wipf,  \pl B251(1989)361. }
\lref \shat {A. Alekseev and S. Shatashvili, \np B323(1989)719.}
\lref \ber  {M. Bershadsky and H. Ooguri, \cmp 126(1989)49.}
\lref \oltu {D. Olive and N. Turok, \np B220(1983)491.}
\lref \klts { C. Klim\v c\'\i k  and A.A. Tseytlin, \pl B323(1994)305;
hep-th/9311012.}
\lref \natd {J.-L.  Gervais, L. O'Raifeartaigh, A.V.  Razumov and M.V.
Saveliev, preprint DIAS-92/27, hep-th/9211088.}

\lref \berg {E. Bergshoeff, I. Entrop and R. Kallosh, preprint SU-ITP-93-37;
hep-th/9401025.}
%%%%%%%%%%%%%%%%%%%%%%%%%%%%%%%%%%%

%%%%%%%%%%%%%%%%%%%%%%%%%%%%%%%%%%%%%

\newsec{Introduction}

 Because of   possible implications of string theory for
gravitational physics it is important  to   study
string solutions  with  realistic signature
($-,+,+,\dots$). In trying to understand   issues of singularities and
short distance structure
one is  mostly interested not just  in  solutions of the leading-order
low-energy  string effective equations but in the ones  that  are exact
in $\a'$ and/or which have  an  explicit  conformal field theory
interpretation.
While the leading-order solutions are  numerous, very few  solutions of the
second type  are known.
 In this contribution,
based on the original work \klitse , we
 present some new    exact   solutions
which correspond to gauged WZNW models and thus  should have a direct conformal
field theory interpretation. The `null' gauging means the gauging of
  WZNW models  \witt\gwzw\gwz\karabali\ based on non-compact
\bal\barn\oraif\   groups
 with the generators of the  gauged subgroup being `null' (having zero Killing
scalar products). The  gauged subgroup  will be  thus chosen  to be solvable
(but  need not be nilpotent in general).
The resulting  \sms will belong to the following class\foot{The so-called
F-model considered in \horotse, where the conditions of (all order in
$\alpha '$) conformal invariance were given for  generic $F$ and $\phi$.}
$$  S = {1\ov \pi \a'} \int d^2 z
 \big[   \del x^i \bd x_i
   +   F(x)   \del u \bd v  \big]   +  {1\ov 4 \pi  }  \int d^2 z \sqrt
{g^{(2)}}  R^{(2)} \p (x)   \  , \eq{0}      $$
where the  two functions $F$ and $\p $  will  be explicitly determined (in
Section 2)  by gauging of  the   nilpotent   subgroups   in WZNW models for
rank
$n$ maximally non-compact groups.  $x^i$ ($i=1,...,n$)  will be the linear
combinations of  the coordinates  $r^i\equiv r^{\a_i}$  corresponding to the
simple roots  $\a_i$,
$$\del x^i \bd x_i=  C_{ij}  \del r^i\bd r^j \ , \ \ \a_i\cdot x = K_{ij} r^j \
, \ \
 \  \ K_{ij}\equiv K_{\a_i\a_j}= {2 \a_i \cdot \a_j \ov  |\a_j|^2}  = \ha
|\a_i|^2  C_{ij} \ , $$
where $ K_{ij}$ is the  $n\times n$ Cartan matrix. We shall  find that
$$ F=  {1\ov   \sum_{i=1}^n \ep_i {\rm e}^{  \a_i\cdot x  } }\ , \  \ \ \ \
\p =  \ha \sum_{s=1}^m  \a_s\cdot x  -  \ha
 \ln   \big(\sum_{i=1}^n  \ep_i {\rm e}^{  \a_i\cdot x }\big) = \r \cdot x   +
\ha \ln F
\   ,   $$
where the constants $\ep_i$ can be chosen to be $ 0$ or $ \pm 1$ and $\r= \ha
\sum_{s=1}^m  \a_s$ is
 half  of the sum of all positive roots.\foot{ If  $\a_1$  is  a simple root
corresponding to the generators $E_{\pm\ao}$
which are left ungauged  (the remaining  $m-1$  positive (negative) roots
correspond to the generators of a  left (right) nilpotent subgroup that was
gauged)
then $\ep_1=1$  ($m= \ha (d-n), \ n= \rank G, \ d=\dim G $).}

\def \ax {  \a_i\cdot x }

In Section 3  we
 discuss an    apparent similarity  to  the Toda model (in particular,  we
shall find a  direct relation between the solutions of the classical equations
of motion).
 Some concluding remarks will be also made , concerning the dual
(in $u+v$ direction) version of the above $\sigma$-model.

%%%%%%%%%%%%%%%%%%%%%%%%%%%%%%%%%%%%%%%%%%%
\newsec{Null gauging of WZNW models}
%%%%%%%%%%%%%%%%%%%%%%%%%%%%%%%%%%%%%%%%%%%%%%%%%%%%%%%%%%%

\subsec{\bf General scheme}

 The indefinite signature of the Killing form for non-compact  algebras implies
that
there is a number of `$null$'  generators  $T_n=N_n$  which have zero scalar
products,
$ \Tr (N_nN_m) = 0$. A subalgebra generated by such generators
is thus solvable (but may not be nilpotent).\foot{An example of  a `null'
 generator in the Lorentz group case
is  a  sum of a spatial rotation with a boost. Note that a nilpotent ($N^2=0$)
generator  is null but, in general, a null generator need not be nilpotent.
Gauging of subgroups generated by nilpotent generators
was previously discussed  in \oraif\dvv\jack\ira\arfa\ind.}
  In this  case  one can  consider  a  left-right asymmetric gauging
since the anomaly cancellation condition \witte\  ($\Tr T_L^2=\Tr T_R^2$)
is   obviously satisfied.

Consider the action of the `null' gauged WZNW model
$$ S = -kI(g)
 -{k\over \pi }
 \int d^2 z \Tr \bigl( - A\,\bd g g\inv +
 \bar A \,g\inv\del g + g\inv A g \bar A   \bigr) =-k I(h^{-1}g\bar h)
  ,   \eq{1} $$
where
$$    I \equiv  {1\over 2\pi }
\int d^2 z  \Tr (\del g^{-1}
\bd g )  +  {i\over  12 \pi   } \int d^3 z \Tr ( g^{-1} dg)^3   \ .
$$
and\foot{$A$ and $\bar A$ should be considered as chiral projections of the
two independent vector fields}
$$A=h \del h\inv,~~~\bar A=\bar h\bd \bar h\inv, ~~~~~h\in H_+, \bar h\in H_-,
$$
Here $H_+,H_-$ are two different subgroups of $G$ generated by null generators
($Tr(N_nN_m)=0$). The action (1) is local and manifestly gauge invariant under
the $H^{\pm}$-gauge transformation
 $$ g \ra w^{-1} g \bu \ , \  \ A \ra w\inv ( A + \del ) w\ , \ \
 \A \ra \bu\inv ( \A + \bd ) \bu \ , \ \  \ \ \ \
 \ \ \ h \ra w\inv h \ , \ \ \bh \ra   \bu\inv \bh  \  ,\eq{2} $$
Assuming that  the  corresponding quantum theory is regularised in the
`left-right decoupled' way (so that  the local
counterterm  $ \Tr (A\A)$   does not appear)
the  only  non-trivial renormalisation that can  occur at the quantum level is
the shift  of the overall coefficient
$k\ra k- \ha c_G$
\foot{$c_G$ is the value of the quadratic Casimir operator in adjoint
representation. The negative sign of the shift is due to our choice of the
`unphysical' sign in the action (1) as usual in the non-compact case.}
 in front of the action (1).  As a result, the  couplings of
the \sm
obtained by integrating out the gauge fields $A,\A$   should  not receive
non-trivial $k\inv$ -corrections,
i.e. they should represent an  exact solution of the  \sm conformal invariance
equations.
The central charge of the resulting gauged model will be equal to the central
charge of the original WZNW model minus the  dimension of the gauged subgroup.
 Depending on a choice of the null subgroups
$H_+$ and $H_-$  the trace of the product of their generators $\Tr (N\bar N)$
 and
hence  $ \Tr (A\bar A)$ may or may not vanish so that (1), in general, is
different
from the action of the vectorially gauged WZNW model
$$ I_v (g,A) = I (h\inv g \bh ) -  I (h\inv \bh)  \
 = I(g)  +{1\over \pi }
 \int d^2 z \Tr \bigl( - A\,\bd g g\inv +
 \bar A \,g\inv\del g  $$ $$ + g\inv A g \bar A  - A \A \bigr)
 \equiv  I_0(g,A) -
 \1p \int
d^2z \Tr ( A\A )  \ ,   \eq{3} $$

%%%%%%%%%%%%%%%%%%%%%%%%%%%%%%%%%%%%%%%%%%%%%%%%%%%%%%%
\subsec{\bf Gauging of nilpotent subgroups  in  Gauss decomposition
parametrisation}
%%%%%%%%%%%%%%%%%%%%%%%%%%%%%%%%%%%%%%%%%%%%%%%%%%%%%
A  particular case of such gauging (when the null subgroups are the nilpotent
subgroups corresponding to  the step  generators in the Gauss decomposition)
was  considered previously \oraif\dvv (see also \jack\ira)  in the context of
Hamiltonian reduction \shat\ber\
of WZNW theories related to  Toda models.\foot {The WZNW model in the Gauss
decomposition parametrisation was considered in \ger. The standard (vector or
axial) gauging  in the  Gauss decomposition  was  also discussed  in \arfa.}
The  approach based on gauging of any  subgroup  with null generators
 is more general since, in principle,  we
do not need  to use  the Gauss decomposition (which does not always exist for
the real groups  we are  to consider to  get  a real WZNW action).
 The gauging based on the  Gauss decomposition   directly applies only to the
groups  with the algebras that are  the `maximally non-compact' real forms of
the classical Lie algebras
(real linear spans of the Cartan-Weyl   basis), i.e.   $sl(n+1,R), \ so(n,n+1),
\ sp(2n,R), \ so(n,n)$. The corresponding WZNW models can be considered as
natural generalisations of the $SL(2,R)$ WZNW model.
  For these groups there exists a real group-valued Gauss
decomposition
$$  g= g_+  g_0 g_- \ , \ \ \  g_+ = \exp ( \sum_{\Phi_+ } u^\a E_{\a }) \ , \
\
g_-= \exp ( \sum_{\Phi_+} v^\a E_{-\a }) \ ,  \eq{4} $$
$$   g_0 = \exp ( \sum_\Delta  r^\a H_\a )=   \exp ( \sum_{i=1}^n  x^i H_i)  \
. $$
Here $\Phi_+$ and $\Delta$ are the sets of the positive and simple roots of a
complex algebra with the Cartan-Weyl  basis consisting of the  step operators
$E_\a, \ E_{-\a}  , \ \a \in \Phi_+ $ and  $n(=\rank G)$  Cartan  subalgebra
generators $ H_\a , \ \a\in \Delta$.\foot{In  what follows we shall assume
that there is always  a sum over  repeated upper and lower indices.
We shall also  use $r^\a$  with understanding   that $r^\a \not=0$ only if $\a$
is a simple root.}
We shall use the following  standard relations  (we  shall assume that a long
root has $|\a|^2 =2$)
\hum\olive\
$$
[ H_\a,E_\b]= K_{\b\a} E_\b \   \  ( \a \in \Delta, \  \b \in \Phi )\  ,
  $$ $$
  [E_\a, E_{-\a} ]= H_\a     \   \ ( \a\in \Delta  ) \    , \ \   \
[ E_\a,E_\b]= N_{\a\b} E_{\a + \b}  \ , $$
$$ \Tr (H_\a H_\b ) \equiv C_{\a\b}={2\ov |\a|^2} K_{\a \b}\ , \ \ \  K_{\a
\b}= {2 \a \cdot \b \ov |\b|^2} \ ,   \   \ \ \Tr (H_i H_j ) =\d_{ij}  \ ,   $$
$$ H_\a = \tilde\a^i H_i  \ , \ \ \
 \ , \ \ \   x^i =   \sum_{\a \in \Delta} \tilde\a^i r^\a \ , \ \  \  K_{\a\b }
r^\b  =\a\cdot x \ , \ \ \ \tilde\a^i\equiv  {2\ov |\a|^2} \a^i \ ,  $$
$$ \Tr (E_\a E_{-\b }) =  {2\ov |\a|^2} \d_{\a \b} \ , \ \   \
 \Tr (E_\a H_\b )=0 \ , \   \eq{5} $$
where  $\a^i \ (i=1,...,n)$ are the components of the positive  root vectors.
It is  clear that $E_\a$ and $E_{-\a}$ form sets of null generators  so that
some
of the corresponding symmetries can be gauged  according to (1). For example,
we   may take $w, \ A $ and $h$ in (2)  to  belong  to the one-dimensional
subgroup generated by some $E_{\g}$    and  $\bu, \A $ and $\bh$ -- to the
subgroup generated by
$E_{-{\g'}} $   where $\g$ and ${\g'}$ are  positive roots which  may not
necessarily be the same.

If one gauges  the full left and right  nilpotent subgroups  $G_+$ and $G_-$
($\dim G_{\pm} = \ha (d-n)= m, \ \dim G = d ) $  generated by
all generators $E_\a$ and $E_{-\a}$ \oraif\ one is left with the action for
$n$  decoupled scalars $r^\a$  or $x^i$
 which represent  the free part of the  Toda model action.
Being interested in finding non-trivial  conformal  \sms
describing string  solutions we  are to consider the more general case of
`partial' gauging  when   only some
subgroups   $H_+$ and $H_-$  of  $G_+$ and $G_-$ are gauged.
  $r^\a$  should correspond to spatial directions ($C_{\a\b}$  in (5) is
positive definite).
Since  the Killing form  of the maximally non-compact groups has $m=\ha (d-n)$
time-like directions,  to get a physical signature  of the  resulting
space-time  we need to gauge
 away all but {\it one}  pair of coordinates $u,v$ in (4). Therefore
 the gauge groups $ H_{\pm}$  should  have dimension $m-1=\ha (d-n)-1$, i.e.
$$ \dim H_\pm = \dim G_\pm - 1  \ . $$
As we shall see below (in Sect.2.3) the ungauged generator(s) of  $G_\pm$
must be a simple root.
Moreover, to get   the physical value    $D=4$  of the target space  dimension
we need to  start with  the  rank 2 groups  $G$  ($D= n + 2=4$).

Let us first consider the most general case   when
 $w, \ A $ and $h$ in (2)   correspond to the subgroup   $H_+ \subset G_+$
generated by some  $s\leq m$
 linear combinations  ${\cal E}_p = \l^\a_p  E_\a $ of the generators  $(E_\a,
\ \a\in \Phi_+ ) $  of $G_+$
and  $\bu, \A $ and $\bh$ -- to the subgroup $H_- \subset G_-$ generated by
some  $s'=s$ linear combinations $\EE_{q}= \ll^\a_q E_{-\a} $.
   Then  it is straightforward to write down the resulting  expression for the
action  (1)  using
the Polyakov-Wiegmann formula  and (5) (i.e.  $I(g_+)=I(g_-)=0$, etc.)
$$ I_n = I (h\inv g \bh) = I(g_0)  +  {1\over \pi }
 \int d^2 z \Tr \bigl[  g_0\inv  g_+\inv h \del (h\inv g_+) g_0  g_-\bh \bd
(\bh\inv g_-\inv)
\bigr]
$$ $$ = I(g_0)  + {1\over \pi }
 \int d^2 z \Tr \bigl[  g_0\inv (A +  g_+\inv \del  g_+) g_0  (\A -  \bd g_-
g_-\inv)
\bigr] \ . \eq{6} $$
Setting
$$   A=  {\cal E}_p B^p = \l^\a_p B^p E_\a\ , \ \ \ \A =   \EE_q  \B^q=
\ll^\a_q \B^qE_{-\a}
 \ ,   $$ $$ \  g_+\inv \del  g_+  \equiv J_u = U^\b_\a (u) \del u^\a  E_\b \ ,
\ \ \  \bd g_-   g_-\inv \equiv \J_v = \J^\a_v E_{-\a} = V ^\b_\a (v) \bd v^\a
E_{-\b}  \ ,  \eq{7} $$
we get
$$ S_n =  {k\ov \pi} \int d^2 z  \big[\ha C_{\a\b} \del r^\a \bd r^\b
 +   M_{\a\b} (J^\a_u  + \l^\a_p B^p) (\J^\b_v - \ll^\b_q \B^q) \big] \  $$
$$ =  {k\ov 2\pi} \int d^2 z  \big[ \del x^i \bd x_i
 +   2M_{\a\b} (J^\a_u  + \l^\a_p B^p) (\J^\b_v - \ll^\b_q \B^q) \big] \   ,
\eq{8}    $$
where
$$ M_{\a\b} = \Tr ( g_0\inv E_\a g_0 E_{-\b} ) = f_\a (r)  \d_{\a\b}  \ , \ \ \
f_\a (r)  \equiv  {2\ov |\a|^2} {\rm e}^{  - K_{\a\b } r^\b } ={2\ov |\a|^2}
{\rm e}^{  - \a\cdot x } \ .  $$
The  sums over $\a,\b$  run  over  positive roots ($r^\a\not=0$   for simple
roots only). It is clear that when  $H_\pm$=$G_\pm$, i.e. when $\l^\a_p$
and $ \ll^\b_q  $  are non-degenerate  we can eliminate $J_u$ and $\J_v$ from
the action by redefining  the gauge fields $B, \B$. One is then left with the
free action for $r^\a$
plus the dilaton term  $\p=  \p_0 +  \ha \sum_\a K_{\a\b } r^\b $
originating from the $B,\B$ -determinant.

Integrating out $B^p$ and $\B^q$  in (8) we get
$$ S_n =  {k\ov \pi} \int d^2 z  \big[\ha C_{\a\b} \del r^\a \bd r^\b
 +   \M_{\a\b} (r)   U^\a_\g (u)  V^\b_\d  (v) \del u^\g  \bd v^\d  \big] $$ $$
  -  {1\ov 8\pi  }  \int d^2 z  \sqrt{ g^{(2)}} R^{(2)} \ln \det  M_{pq} (r)  \
  \  ,  \eq{9}  $$
$$  M_{pq}(r) \equiv   M_{\a\b}  \l^\a_p  \ll^\b_q = \sum_\a  f_\a (r)  \l^\a_p
 \ll^\a_q \ , \ \
 \   $$ $$  \M_{\a\b} (r) =M_{\a\b} - M^{-1 pq} \ll^\g_p \l^\d_q
M_{\a\g}M_{\b\d} =
 f_\a   \d_{\a\b} -   f_\a f_\b M^{-1 pq} \ll_{\a p } \l_{\b q}   \ .  \eq{10}
$$

%%%%%%%%%%%%%%%%%%%%%%%%%%%%%%%%%%%%%%%
\subsec{\bf Models with one time-like coordinate}
%%%%%%%%%%%%%%%%%%%%%%%%%%%%%%%%%%%%%%%%%%

Let us now  turn  to the most interesting case
when  the dimensions of the gauge groups $H_\pm$ are equal to $\dim G_\pm
-1=m-1$
so that only one time-like  coordinate appears in the  resulting \sm action.
Let  $E_\ao$  and $E_\am $  denote the generators of $G_+$  and $G_-$  which
remain ungauged, i.e.
which do not belong to $H_+$ and $H_-$. Since $H_+$  must  be a subgroup,
 $E_\ao$  cannot appear in the commutators of the generators
of $H_+$.  According to (5) this is possible only if  $\ao$ is a simple root,
i.e.  if it cannot be represented as a sum of two other positive roots.
In fact, if  we use the indices $i,j$ to denote the simple roots $\a=\a_i \
(i=(1,s)=1,2,...,n)$   and indices $a,b$ to denote the remaining positive roots
$\a_a \ (a=n+1,...,m)$  the commutators of the corresponding  step operators
are given by
$$ [E_i,E_j] \sim E_a \  \  (\a_a= \a_i + \a_j) \  ,  $$ $$  [E_i, E_a] \sim
E_b \ \   (\a_b= \a_i + \a_a)\ , \  \  \  [E_a, E_b] \sim E_c \  \  (\a_c= \a_a
+ \a_b) \   . $$
It is  clear that one can also use  linear combinations $E_s'= E_s + \l_s E_1 \
(s=2,...,n)$
as the  `simple' part of the generators of $H_+$ but one cannot mix the
non-simple generators $E_a$ with $E_1$.

Let $ u^\ao \equiv  {1\ov \sqrt 2}u, \ v^\ao \equiv {1\ov \sqrt 2}  v$; the
remaining coordinates $u^\s, v^\s$ ($\s=(s,a)$  will be used to denote  all
`gauged'  $m-1$  positive roots)
 are transforming under the gauge group (with the leading-order term being just
a shift)
so that we can set them to zero as a gauge. In this gauge $J_u =  {1\ov \sqrt
2}\del u E_\ao, \
\J_v =  {1\ov \sqrt 2} \bd  v E_\am$ and the \sm  action  (9) takes the  form
($p,q=1,...,m-1$)\foot{For notational convenience (to get rid of an extra
factor of 2 in front of the $  F (x)   \del u  \bd v  $ term)  we have
redefined $u$ and $v$ by the factor of $1/\sqrt 2$ as compared to (9). }
$$ S_n =  {k\ov 2\pi} \int d^2 z  \big[ \del x^i \bd x_i
 +   F (x)   \del u  \bd v  \big]    +  {1\ov 4\pi  }  \int d^2 z  \sqrt{
g^{(2)}} R^{(2)}  \p (x)  \ ,   \eq{11 }  $$
$$  F(x) = f_\ao   -   f^2_\ao  M^{-1 pq} \ll^\ao_p \l^\ao_q  \ , \ \ \p (x) =
- \ha \ln \det  M_{pq} \ , $$ $$ \ \ M_{pq}(x) = \sum_\a  f_\a (x)  \l^\a_p
\ll^\a_q \ . $$
The non-trivial elements of the `mixing' matrix  $ \l^\a_p  $
correspond to a possibility of changing the generators of $H_+$ by adding
$\l^\ao_p E_\ao $.
Without loss   of  generality   the non-vanishing components of $\l^\a_p$ can
be taken  to be:
 $ \l^\s_p = \d^\s_p , \ \  \l^\ao_p = \l_s \d_{ps}$     and similarly for
$\ll^\a_q$ (according to the remark above, only simple roots can be mixed with
$E_\ao$).
 Then  ($s,t=2,...,n; \ a,b= n+1,...,m$)
$$  M_{pq}=    \left(\matrix
{M_{st} &0\cr 0&M_{ab}\cr } \right) , \ \
 M_{st} (r) = f_s \d_{st}   +   f_1  \l_s  \ll_t  \ , \ \ M_{ab} (x) =  f_a (x)
\d_{ab}\ , \  \eq{12}  $$
$$
f_h (x) \equiv f_{\a_h} = {2\ov |\a_h|^2} {\rm e}^{  - \a_h \cdot x   }  \ , \
\ \   \ h=(1,s,a)=1,2,...,m \ , \ \ m
=\ha (d-n)\  .
 \eq{13} $$
If we  introduce  $\l_1=\ll_1 =1 $  in order to make the formulas look
symmetric with respect to all simple roots,
we find
$$ M_{st}\inv =  f_s\inv \d_{st}   -
    { f_s\inv f_t\inv \l_s  \ll_t \ov  \sum_{i =1}^n f\inv_{i} \l_i \ll_i} \ ,
\ \ \  \det M_{pq} =  \big(\prod_{h=1}^m f_h \big) \big(\sum_{i =1}^n f\inv_i
\l_i \ll_i  \big)\ . \eq{14}   $$
As a result,
$$    F  = f_1   -   f_1^2  M^{-1 st} \ll_s \l_t  = { 1 \ov
 \sum_{i =1}^n f\inv_i \l_i \ll_i }  \ , $$ $$  F  =
 \bigl( \sum_{i =1}^n  \ep_i  {\rm e}^{  \a_i \cdot x  } \bigr)\inv \ ,  \ \ \
\  \  \ \ep_i\equiv
 \ha {|\a_i|^2} \l_i \ll_i \ ,   \eq{15} $$
$$
 \p= - \ha \sum_{h=1}^m \ln  f_h  - \ha  \ln  \sum_{i =1}^n f\inv_i \l_i \ll_i
\ ,  $$ $$
\p =\p_0   +  \r \cdot x   -
\ha  \ln   \sum_{i =1}^n  \ep_i  {\rm e}^{  \ax  }  \ , \ \ \ \  \r= \ha
\sum_{h=1}^m  \a_h \ .
    \eq{16} $$

We   have  thus found   the \sm action  (11),(15),(16), i.e.\foot{We absorb the
prefactors $
2/|\a_i|^2 $ of  the exponential  terms into a rescaling of $u,v$ and $\ep_i$.
We also consider (17) as an effective action, including the  quantum shift of
$k$,
 $ \ \k\equiv k- \ha c_G$. }
$$ S_n =  {\k\ov 2\pi} \int d^2 z  \big[ \del x^i \bd x_i
 +      \bigl( \sum_{i =1}^n  \ep_i  {\rm e}^{  \a_i\cdot x  } \bigr)\inv  \del
u  \bd v  \big]    $$
$$ +  {1\ov 4\pi  }  \int d^2 z  \sqrt{ g^{(2)}} R^{(2)}  \big( \r\cdot x    -
\ha
  \ln   \sum_{i =1}^n  \ep_i  {\rm e}^{  \a_i\cdot x   } \big)   \ ,    \eq{17
}  $$
with $\a_i$ being the simple roots and $\r$ being  half the sum of all positive
roots.  The  values of the  parameters
$\ep_i=0, +1$ or $ -1$
  represent  inequivalent   gaugings of the original WZNW model or different
conformal sigma models.\foot{
Inequivalent solutions  corresponding to different possible choices of
an   ungauged simple root $\a_1$  are easily included by assuming that $\ep_1$
can also take values $0$ and $-1$ but at least one of $\ep_i$ is non-vanishing.
 In general,   $\ep_i$  taking arbitrary real values  represent  moduli of the
solutions.}  Non-trivial models (not equivalent to direct products of $SL(2,R)
$ WZNW with free scalars) are found for non-vanishing values of the  `mixing'
parameters $\ep_2, ...,\ep_n$.

 The  metric of the corresponding $D=n+2$ dimensional target space-time has two
null Killing  symmetries (in fact, the full $2d$ Poincare  invariance in the
$u,v$ plane, $\ u'= \r u + a, \ v' = \r\inv v + b$).
The    non-trivial $(uv)$  components of the metric and  the antisymmetric
tensor
and the non-linear part of the dilaton  are all  expressed in terms of a
single function   $F(x)$  (15),
which is the inverse of the sum of  the exponentials of the spatial Cartan
coordinates $x^i$. The metric is non-singular if all $\ep_i$ have the same
sign.
\newsec{Relation to the Toda models}

To  demonstrate the  equivalence  to the Toda model at the classical  level
 let us consider the
classical equations for the model  (0) (on a flat $2d$ background)
$$  \del\bd x_i  -\ha \del_i F \del u \bd v =0 \ , \ \ \
\del( F \bd v) =0 \ , \ \ \ \bd (F \del u) =0 \ .  \eq{18} $$
The model thus has two chiral currents. Integrating the last two equations and
substituting the
solutions in the first one  we get
$$ \del\bd x_i  + \ha  \chi \del_i F\inv =0 \ , \ \ \  F \bd v = \n (\bar z)  \
, \  \ F \del u = \m (z)
\ , \  \  \chi \equiv
\n (\bar z) \m (z) \  . \eq{19} $$
Since $\n, \m$ are chiral and $\chi$  has  a factorised form  they can   be
made constant by  the
conformal transformations of $z$ and $\bar z$ (the  \sm  (0) is always
conformally invariant at the classical level).
Equivalently, this can be  considered as a  gauge choice (alternative to the
light cone gauge), i.e. $u= a(\tau + \s), \ v= b(\tau-\sigma)$,   for the
conformal symmetry.
The  equation  for $x^i$ can be derived from the action
$$  S = {1\ov \pi \a'} \int d^2 z
 \big[   \del x^i \bd x_i
    -  \chi  F\inv  (x)  \big]  \  .    \eq{ 20 } $$
With $F$ given by   (15) (i.e. $F\inv = T= \sum_{i=1}^n \ep_i {\rm e}^{  \a_i
\cdot x }$)
 and  in the conformal gauge with   constant $\chi$
the  equation  for $x^i$ (19) and the action  (20) are  exactly
those of the Toda model.
This observation is, of course,  related to the derivation of the Toda model
from constrained WZNW model in \oraif.

As a consequence, the equations of the classical string propagation (including
the constraints)
on the backgrounds  (17) discussed  in the present  paper   are    exactly
integrable  since  their solutions can be directly expressed  in terms of
 the Toda model
solutions.

Another   interesting  property of the model (0)  is that  the dual model
obtained by the standard  abelian
 duality transformation  in the $w= \ha (u+v)$  direction
has a  covariantly constant null Killing vector, i.e.
has   a  `plane wave' - type  structure
  (with the metric and dilaton  depending only  on  transverse  coordinates but
not on a  light-cone coordinate).
This is  a consequence of  the fact that the $uv$-component of the metric in
(0) is equal to the corresponding component of the antisymmetric tensor (what
is also the reason for the existence of the two chiral currents).
 Following the standard steps \busc\giveon\
of gauging the symmetry $u'=u + a, \ v'= v + a$,  i.e. adding the gauge field
strength term with a  Lagrange multiplier $ \tu$
and integrating out the gauge field,
we find for the dual model ($t= \ha (u-v)\equiv \tv$)
 $$  \tilde S = {1\ov \pi \a'} \int d^2 z
 \big[   \del x^i \bd x_i   + \TF(x) \del \tu \bd \tu   -  2 \del \tv \bd \tu
   \big]   +  {1\ov 4 \pi  }  \int d^2 z \sqrt {g^{(2)}}  R^{(2)} \tp (x) \   ,
   \eq{21} $$
$$ \TF= F\inv (x) \ , \ \ \ \ \ \tp = \p (x)  -  \ha \ln F (x)   \ . \eq{22} $$
The dual  space-time  metric   has one  covariantly constant null
Killing  vector while  the   antisymmetric tensor vanishes.
Since the `transverse' part of the metric is flat,
the conditions of  the  Weyl invariance of the  model  (21)  are  given  (to
all orders in $\a'$)
  by  the `one-loop' conditions\foot{See,   e.g.,
\tsnul\ for a  general discussion.  Since the dilaton is assumed to depend on
the  transverse coordinates,
this model  is a  generalisation of the original `plane-wave' models
considered in \guv\amkl\hst\rudd.  Some special  cases, in particular,  the
$D=3$  case   of such model
 -- the duality rotation  of the $SL(2,R)$ WZNW model -- were   already
discussed
  (in connection with extremal black strings) in \horhorst\horwel.}
$$ -   \del^i \del_i   \TF  +  2\del_i \tp \del^i \TF
   = 0  \ , \ \ \  \del_i \del_j   \tp = 0 \ ,  \eq{23} $$
$$   {26-D \ov 6 \a' }=  - \ha  \del_i \del^i
  \tp   +  \del_i \tp \del^i \tp  \
   .     \   $$
The dual data to (15) obviously solve those equations, thus providing
 another example  of exact solutions
related by the standard (leading-order) duality (cf. \klts).
The classical equations of motion of the two dual models are, of course,  also
equivalent and  can be represented in the Toda-like  form
(19).

%%%%%

\vfill\eject
\listrefs

\end